\documentclass[prl,twocolumn,floatfix,nobibnotes,superscriptaddress]{revtex4}
\usepackage{amsmath}
\usepackage{graphicx}
\usepackage[tight]{subfigure}
\usepackage{Jonasmacros}

\begin{document}
\title{Disorder induced metallicity in amorphous graphene}
\author{Erik Holmstr\"om}
\affiliation{Instituto de Ciencias F\'isicas y Matem\'aticas,
Universidad Austral de Chile, Valdivia, Chile}
\author{Jonas Fransson}
\affiliation{Department of Physics and Astronomy, Division of Materials Theory, Uppsala University,
Box 516, SE-751210, Uppsala, Sweden}
\author{Olle Eriksson}
\affiliation{Department of Physics and Astronomy, Division of Materials Theory, Uppsala University,
Box 516, SE-751210, Uppsala, Sweden}
\author{Raquel Liz\'arraga}
\affiliation{Instituto de Ciencias F\'isicas y Matem\'aticas,
Universidad Austral de Chile, Valdivia, Chile}
\author{Biplab Sanyal}
\affiliation{Department of Physics and Astronomy, Division of Materials Theory, Uppsala University,
Box 516, SE-751210, Uppsala, Sweden}
\date{\today }
\author{Mikhail I. Katsnelson}
\affiliation{Radboud University Nijmegen, Institute for Molecules
and Materials, NL-6525 AJ Nijmegen, The Netherlands}

\begin{abstract}
We predict a transition to metallicity when a sufficient amount of disorder is
induced in graphene. Calculations were performed by means of a first principles
stochastic quench method. The resulting amorphous graphene can be seen as
nanopatches of graphene that are connected by a network of disordered small and
large carbon rings.  The buckling is minimal and we believe that it is a result
of averaging of counteracting random in-plane stress forces.  The linear
response conductance is obtained by a model theory as function of lattice
distortions.  Such metallic behaviour is a much desired property for
functionalisation of graphene to realize a transparent conductor, e.g. suitable for
touch-screen devices.
\end{abstract}
\pacs{later}

\maketitle

Since the discovery of graphene, and all its unique physical
properties \cite{first,r1,r2,r3,r4}, attention is now turned to
functionalization of this material to suit specific applications.
One example is the suggestion of graphane, where H atoms are
adsorbed on graphene. Graphane, which is an insulator, was
predicted from first principles theory \cite{graphane1}, and
was subsequently realized experimentally \cite{elias}. Further
theoretical works \cite{graphane2,graphane3} addressed e.g. the value
of the band gap, which was found to be of order 5.7 eV. In a
similar fashion, fluorine can also be found to adsorb, and in
theoretical works a band gap of 7.4 eV is found
\cite{graphxene1,belgian}, which is larger than the measured value
of 3.4 eV \cite{graphxene2}.  The reason that a large band gap
opens up when hydrogen or fluorine is adsorbed on graphene is that
sp$^2$ bonded C atoms become sp$^3$ bonded. Hence it seems
possible to decrease the conductivity by chemical
functionalization, and turn the semi-metallic graphene to a
semi-conductor or even an insulator. Smaller values of the gap can be
reached by a functionalization by organic molecules \cite{BK2010}.

The ways to increase the conductivity by chemical means has also
been discussed, although here significantly less success can be
identified. Adsorption by single impurities \cite{wehling} have
been explored, as well as the replacement of C atoms for B or N
atoms \cite{biplab}, or even the creation of structural defects in
the C matrix \cite{victoria,jafri,karel}. The electronic structure
of non-crystalline graphene, e.g. around grain boundaries have
also been under consideration, both from tight-binding analysis
\cite{malola} and within the framework of first principles theory
\cite{louie} in combination with calculations of transport
properties. In the works mentioned above, the conducting
properties are still those of a doped semiconductor, with
relatively few charge carriers. A fully metallic behavior was
however suggested in an amorphous structure of graphene
\cite{kapko}, where Stone-Wales defects were introduced into
graphene and geometry optimization was done according to a
Keating-like potential. After geometry optimization, the
electronic structure was calculated from a tight-binding
Hamiltonian, and a non-zero value of the density of states (DOS)
at the Fermi level (E$_F$) was obtained, suggesting that metallic
conductivity is possible. The possibility of a transparent,
mono-atomic thin material, has a great potential in applications
involving touch screens and electronics \cite{ITO}. Very recently,
the two-dimensional amorphous carbon has been derived
experimentally by electron beam irradiation \cite{Kotakoski2011}.
In this paper we study its electronic structure with a
first-principles based approach, which accurately described both
the chemical bonding and structural properties, as well as the
electron energy spectrum.


The amorphous graphene structures were obtained by means of a
stochastic quenching method
\cite{Holmstrom2009b,Holmstrom2010a}. The method is well adapted
to describe amorphous structures that have been obtained by
ultra-fast cooling, such as e.g. sputtered amorphous films where
the structure has been realized directly from the gas phase
\cite{Arhammar2011}. It can also be used to describe the most
typical amorphous structures of a material in order to find
reliable parameters for a Hamiltonian description of the liquid
state \cite{Bock2010}. Normally, this technique is used for
calculations of amorphous structures in bulk where the atoms are
first placed randomly in the calculation cell and then the
positions are relaxed by means of a conjugate gradient method
until the force on every atom is zero. In the present case we
confined the atoms to the plane in order to find a reliable
description of the two-dimensional amorphous structure. First, 200
atoms were placed at random positions in the plane with an average
density equal to that of graphene. We then relaxed the positions
of all atoms, while enforcing that all atoms should stay
in-plane.  Then the area of the plane was relaxed until the
pressure upon the cell was zero.
At this point we obtained the structure called "basic planar". The
area relaxation resulted in new forces on the atoms so we had to
repeat the position and area relaxations until both the pressure
and forces were minimized.
At this point we obtained the structure called "final planar". The
relaxation of positions was then allowed to take place in 3
dimensions and the process of area and position relaxation cycling
was repeated until the in-plane stress and all forces on the atoms
were minimized again.
At this point we obtained the structure called "buckled".

The first principles, self-consistent electronic structure
calculations were performed by means of density functional theory
\cite{Kohn65} using the projected augmented wave method
\cite{Blochl94} as implemented in the Vienna ab initio simulation
package (VASP) code \cite{Kresse96}.  
The geometry optimizations have been performed without any
symmetry constraint and using the conjugated gradient algorithm.
A plane wave energy cutoff
of 300 eV was used for the structural relaxations whereas 500 eV
was used for the DOS and total energy calculations.  The
Perdew-Wang \cite{PW91} parametrization of the generalized
gradient approximation of the exchange-correlation interaction was
used.
For sampling the irreducible Brillouin zone the $\Gamma$ point was used for
the relaxations whereas a 5x5x1 mesh of 13 k-points was used for
the DOS and total energy calculations.  The amount of vacuum in
the supercell was chosen to avoid artificial surface-surface
interactions (24 \AA). In all relaxations, convergence was achieved
when the force on each ion was less than 10$^{-2}$ eV/\AA.

\begin{figure}
\centering
\subfigure[]{%
\includegraphics*[width=.29\columnwidth]{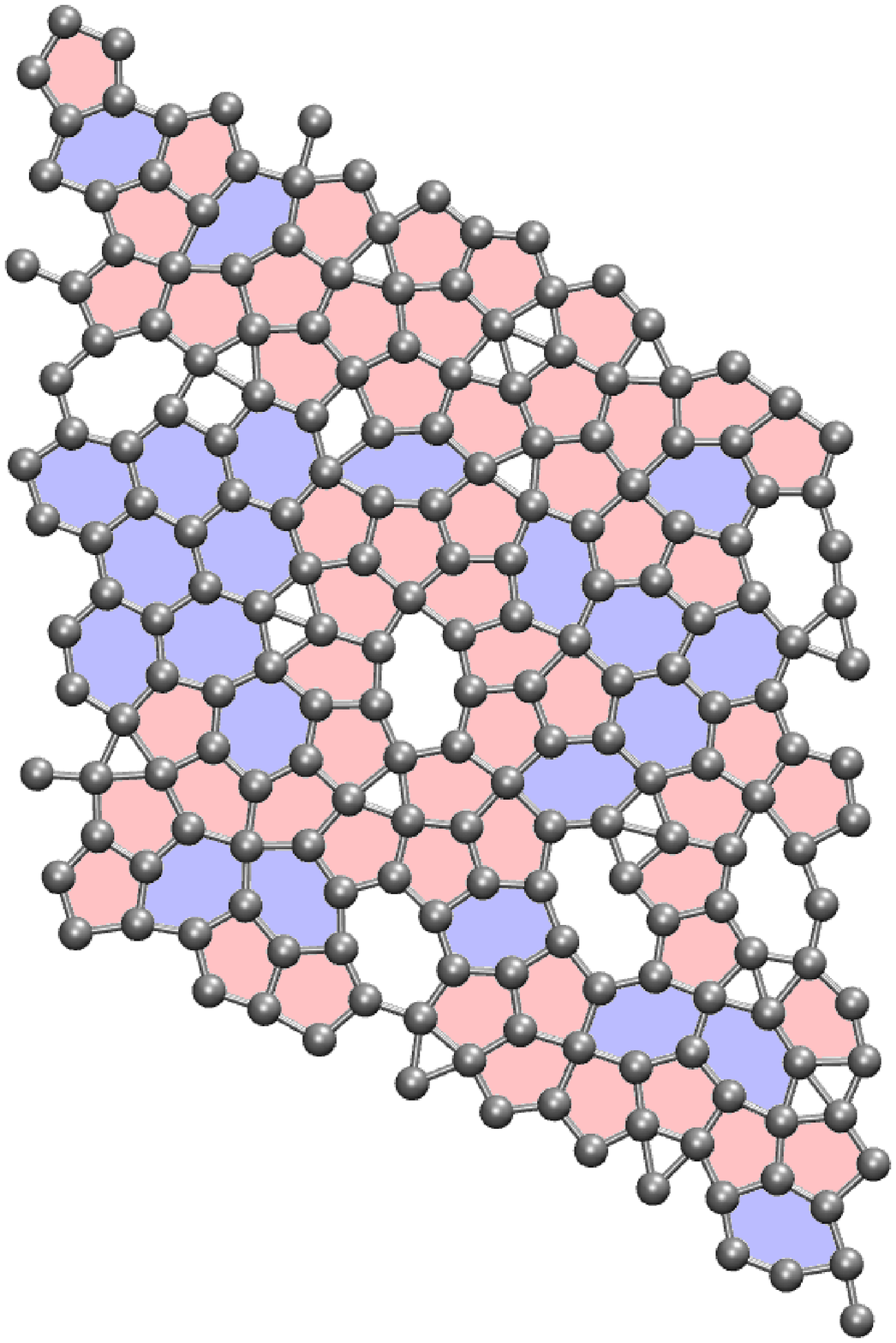}%
}%
\subfigure[]{%
\includegraphics*[width=.29\columnwidth]{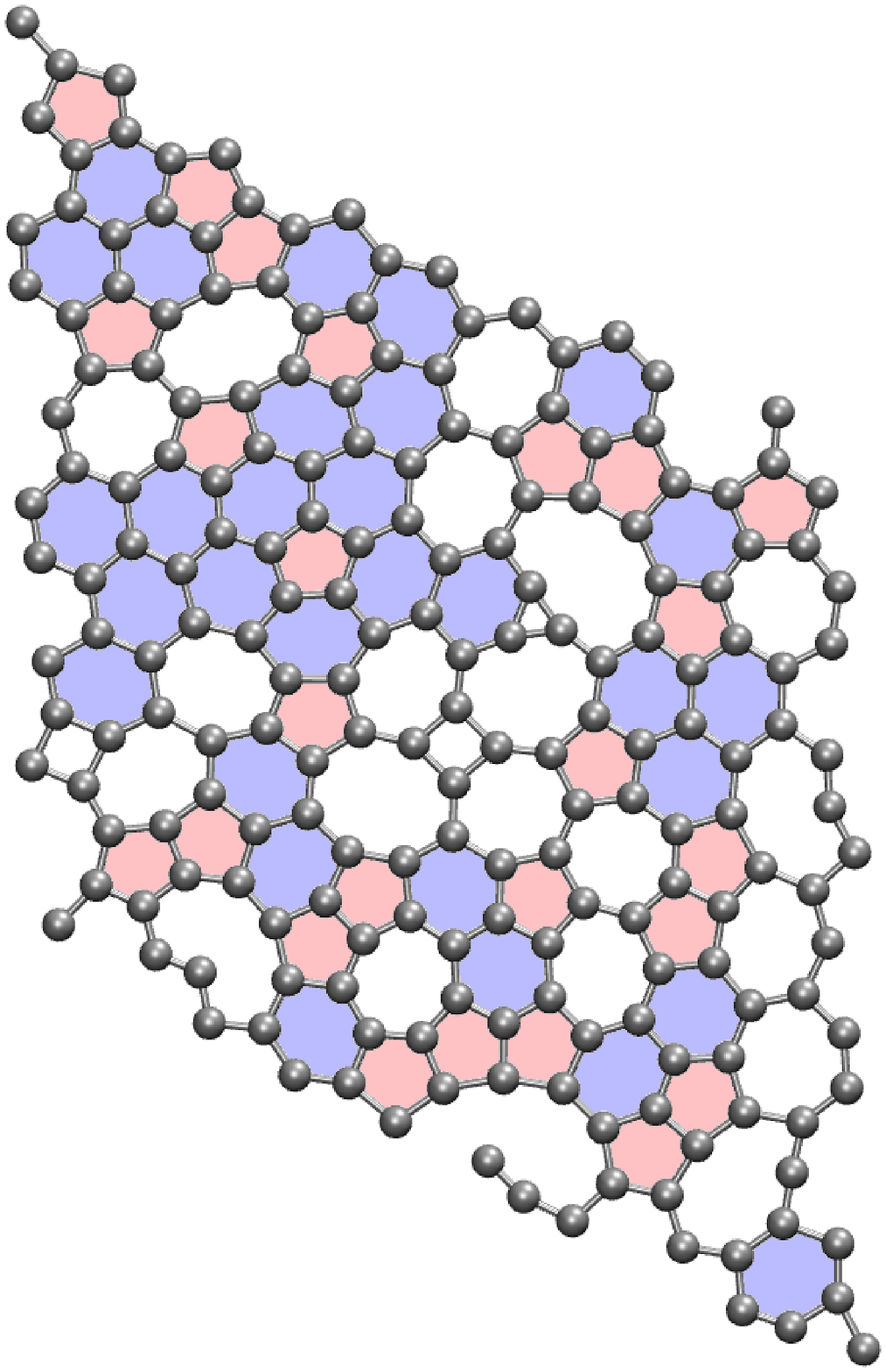}%
}%
\subfigure[]{%
\includegraphics*[width=.29\columnwidth]{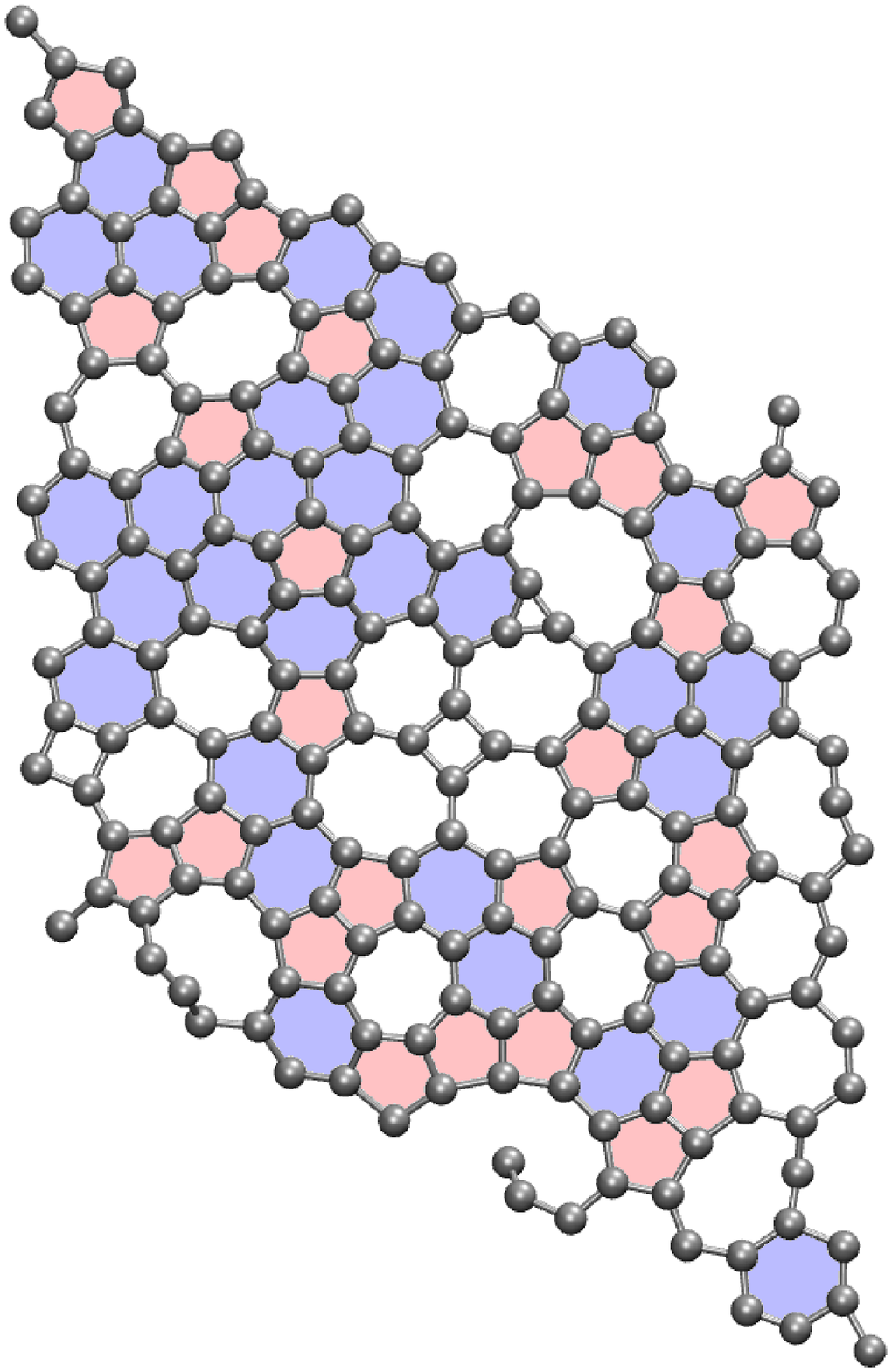}%
}\\
\subfigure[]{%
\includegraphics*[width=.8\columnwidth]{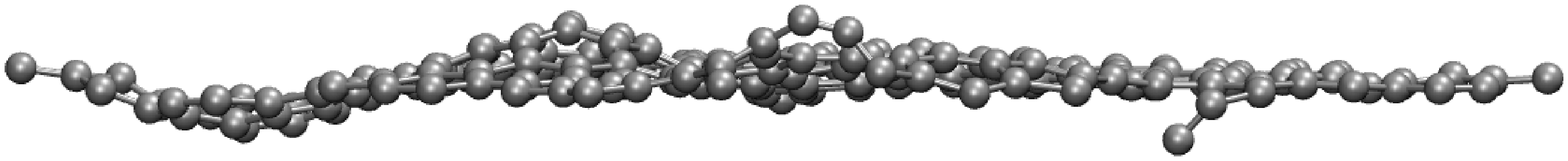}%
}%
\centering%
\caption{(Color online)
(a) Top view of the basic planar structure. Hexagonal rings are marked
with blue and pentagonal rings are marked red.
(b) Top view of the final planar geometry.
(c) Top view of the buckled geometry.
(d) Side view of the buckled geometry. The distance from
left to right in (d) is 41 \AA~ and the buckling amplitude is
about 1.7 \AA. 
}%
\label{fig:structures}%
\end{figure}

The basic planar configuration that was obtained after the first
position-area relaxation cycle is shown
in Figure \ref{fig:structures}(a). We can see that
this structure consists of 23 hexagonal and 56
pentagonal, some tetragonal and triangular rings.  There are also
some areas with larger rings of up to 8 atoms. This structure is
still very compressed since the area is still close to that of
graphene. We can see that a large number of carbon atoms are both
4 and 5-fold coordinated, which indicates a frustrated state of
high energy. The total energy per atom of this structure is 1.50
eV larger than the graphene reference. Further iterations of
positions and area relaxations resulted in the final planar
structure shown in Figure \ref{fig:structures}(b). The
number of hexagonal  rings has now increased to 38 and the number
of pentagonal rings decreased to 25. Several larger rings have
appeared with up to 9 atoms. All carbon atoms are now 3-fold
coordinated and hence sp$^2$ bonded, except in 2 cases where the
coordination is 2-fold with a 180$^{\circ}$ bond angle and hence
the bonding is sp. There is also one atom in the lower edge of the
figure that is bonded to only one other atom. The total energy per
atom of this structure is now 0.50 eV larger than the graphene
reference.

The position and area relaxation of the final planar structure was taken further
by letting the atoms relax also in the direction normal to the plane. The
resulting buckled structure is shown in Figure \ref{fig:structures}(c)
and a side view can be seen in Figure \ref{fig:structures}(d). From the side view we
can see that the plane has evolved into a weakly oscillating shape. In this
structure the number of pentagonal and hexagonal rings has not changed compared
to the final planar structure. The atom with only one bond is also present in
this structure and it can be seen in Figure \ref{fig:structures}(d) where it
protrudes out below the surface. The total energy of the buckled system was 0.43
eV larger per atom than the total energy of the graphene reference. This
structure must be considered as an archetype structure amonst many possible realizations of amorphous
graphene. There may even exist lower energy amorphous states than the one we
have found here. The amorphous state that we have found is surprisingly flat and
contains both squares and triangles. 
We speculate that the flatness is due to an averaging of defect-induced buckling forces.
Square geometries have recently been observed in defect rich graphene that was created
by electron irradiation \cite{Kotakoski2011}.

\begin{figure}
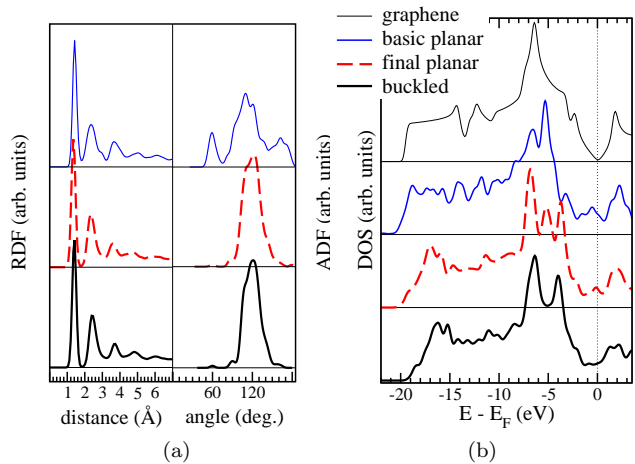

\centering
\subfigure[ ]{%
\includegraphics*[width=0.502\columnwidth]{RDF_ADF_New_4.eps}
\label{fig:RDF}
}\hspace{-0.3cm}%
\subfigure[ ]{
\includegraphics*[width=0.47\columnwidth]{DOS_New_5.eps}
\label{fig:DOS}
}
\caption{(Color online) (a) The radial distribution functions (RDF) for all structures together
           with the angle distribution functions (ADF). (b) Density of states of the
           three amorphous geometries compared to the graphene
           structure.
         }

\end{figure}

The radial and angle 
distribution functions are shown in Figure
\ref{fig:RDF}. We can see that the first peak in all radial
distribution functions is located at 1.4 \AA. However, the radial
distribution function of the basic planar structure has somewhat
broader peaks than the other structures.  The angle distribution
functions show a much larger difference. Here the basic planar
structure shows three pronounced peaks around 60, 110, and 160
degrees whereas the final planar and buckled structures show one
pronounced peak around 120 degrees and only small quantities of 60
and 90 degree angles.

The total DOS for graphene and the three
amorphous geometries are shown in Figure \ref{fig:DOS}.  We can
see that the basic planar structure has somewhat broader bands but
the widths of the final planar and buckled structure are somewhat
narrower  than graphene. The largest difference is however
around the Fermi level where the amorphous structures have high enough
density of states to be clearly metallic.


The realization of a metallic DOS for a monolayer-thick graphene
material is a significant result, and the transport properties of
this system requires further analysis. We do this by a model
theory, as described below. We consider the case of an imperfect
2D carbon lattice where all C atoms are interconnected to three
other C atoms and consider it to be bipartite. In the spirit of a nearest neighbor model, we have
as a Hamiltonian of the system \cite{jafri}
\begin{align}
\Hamil=&
    \sum_{\bfk\bfk'}\tilde{\phi}(\bfk,\bfk')a^\dagger_\bfk b_{\bfk'}+H.c.,
\end{align}
where
$\tilde{\phi}(\bfk,\bfk')=-(t/N)\sum_{\av{ij}}e^{-i\bfk\cdot\bfr_i+i\bfk'\cdot\bfr_j}$,
operators $a$ and $b$ annihilate electrons in sublattices A and B,
respectively. At each site $i$, we express the spatial vectors to the
neighboring sites $\bfr_j$, can be expressed in terms of $\bfr_i$
and additional piece $\bfdelta_{ij}$, i.e.
$\bfr_j=\bfr_i+\bfdelta_{ij}$. Assuming that the random vector
$\bfdelta_{ij}$ connecting sites $i$ and $j$ are independent of
the specific location in the lattice, we can simplify the random
vector to depend only on $j$. This allows the calculation
$\tilde{\phi}(\bfk,\bfk')=-(t/N)\sum_{\av{ij}}e^{-i(\bfk-\bfk')\cdot\bfr_i+i\bfk'\cdot\bfdelta_j}=-t\delta(\bfk-\bfk')\sum_{j=1,2,3}e^{i\bfk\cdot\bfdelta_j}$,
so that we can define
$\phi(\bfk)=\delta(\bfk-\bfk')\tilde{\phi}(\bfk,\bfk')$. The
Hamiltonian is thereby reduced to
\begin{align}
\Hamil=&
    \sum_\bfk\phi(\bfk)a^\dagger_\bfk b_\bfk+H.c.
\label{eq-modelHam}
\end{align}

Next, we assume that the random vector $\bfdelta_j$, $j=1,2,3$, is
normal distributed with mean $\bfm_j=(m_{jx},m_{jy})$ and variance
$\bfsigma_j=(\sigma_{jx},\sigma_{jy})$. Also, by assuming that the
random variables $\delta_{jx}$ and $\delta_{jy}$ are independent, we
can replace the exponential $e^{i\bfk\cdot\bfdelta_j}$ by
$e^{i\bfk\cdot\bfm_j-\sigma_{jx}^2k_x^2/2-\sigma_{jy}^2k_y^2/2}$.
As mean values of the random vectors, we take a set of basic
lattice vectors for graphene, e.g. $\bfm_1=a(\sqrt{3},1)/2$,
$\bfm_2=a(-\sqrt{3},1)/2$, and $\bfm_3=a(0,-1)$, and we assume
that $\sigma_{jx}=\sigma_{jy}=\sigma_j$. For structures with only
small variations from the perfect graphene lattice, we can expand
around the $K$ points $\pm\bfK=\pm2\pi(\sqrt{3}/3,1)/3a$, around
which the dispersion relation for graphene is linear, i.e.
$E_\text{gr}(\bfk)=|\phi_\text{gr}(\bfk\pm\bfK)|=|\pm v_Fke^{\pm
i(\pi/3-\varphi)}|=v_Fk$, where $v_F=3at/2$ and
$\tan\varphi=k_y/k_x$. Expanding the potential $\phi(\bfk)$ around
$\pm\bfK$, retaining contributions of quadratic order in $k$ at most,
we obtain
\begin{align}
\phi(\bfk\pm\bfK)\approx&
        \pm\tilde{v}k
    \biggl(
        1\mp\alpha k
    \biggr)
    e^{\pm i(\pi/3-\varphi)}
\label{eq-disp}
\end{align}
where we have introduced
$\alpha=(2\pi\sqrt{3}\sigma^2/9a)\cos(\varphi-\pi/3)$ and
$\tilde{v}=v_F\exp\{-6(2\pi\sigma/9a)^2\}$, whereas
$\sigma^2=\sum_j\sigma_j^2$. The scattering potential converges to
the usual graphene potential as the variance
$\sigma^2\rightarrow0$, as required. The above expression for the
scattering potential holds for small distortions from the ideal
graphene lattice. Quantitatively we require that
$\sigma\ll\sigma_u=\sqrt{3\sqrt{3}a/4\pi k_c}$, where
$k_c=\sqrt{8\pi/3\sqrt{3}a^2}$ is a high energy cut-off, giving
$\sigma_u=3a\sqrt[4]{\sqrt{3}/2\pi^3}/2\approx0.4a$.

\begin{figure}
 \begin{center}
 \includegraphics*[width=0.99\columnwidth]{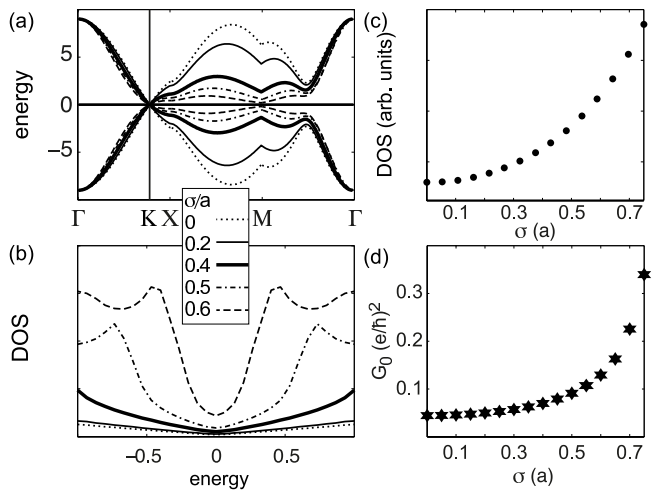}
 \end{center}
 \caption{(a), (b) Band diagram and DOS calculated within the
          stochastic model based on Eq. (\ref{eq-modelHam}) for varying degree of lattice
          distortion, (c) DOS($E_F$) calculated using the approximate dispersion relation
          in Eq. (\ref{eq-disp}), and (d) linear response conductance, at low
          temperatures, based on the calculations in (a) and (b). The residual DOS and
          conductance at zero lattice distortion in panels (c) and (d) is due to an
          artificial finite broadening ($\eta=0.05$) introduced in the Green function.}
 \label{fig-model}
 \end{figure}

Numerically, we find that the dispersion given in Eq.  (\ref{eq-disp}) is
applicable for lattice distortions up to about $\sigma/a\lesssim0.5$, see Fig.
\ref{fig-model} (a), where we plot the computed bands when applying the
stochastic arguments to the model given in Eq. (\ref{eq-modelHam}). Around $K$,
the dispersion is basically linear for rather large lattice distortions, and the
DOS resulting from the band structure, Fig.  \ref{fig-model} (b), shows the
typical linear characteristics around the Fermi level ($E_F=0$) for small
lattice distortions ($\sigma/a\lesssim0.5$). Here, the DOS ($\propto\im{G^r}$)
is calculated in terms of the Green function $G^r(\bfk,\omega)={\bf
N}^{-1}[\varphi^\dagger(\bfk)\varphi(\bfk)/(\omega-{\cal E}(\bfk)+i\eta)]$ is
defined in terms of the eigenbasis $\{\varphi(\bfk),{\cal E}(\bfk)\}$ given by
the Hamiltonian in Eq. (\ref{eq-modelHam}).
In Fig. \ref{fig-model} (c), we plot the DOS resulting from the
approximate dispersion in Eq. (\ref{eq-disp}), showing a monotonic
increase with the lattice distortion, as expected.

For further investigations of the potential metallicity of the
amporphous graphene, we also calculate the linear response
conductance, $G_0$, using the Kubo formalism. For an electric field applied in the $x$-direction, we use
\begin{align}
G_0=&
	\lim_{\omega\rightarrow0}\frac{1}{\omega}
	\int_0^{\infty}
		\av{\com{j_x^\dagger(\bfq,\tau)}{j_x(\bfq,t)}}e^{i\omega\tau}d\tau,
\end{align}
where $\lim_{\bfq\rightarrow0}j_x(\bfq,\tau)=ie\sum_\bfk\Phi(\bfk)a^\dagger_\bfk
b_\bfk+H.c.$, and $\Phi(\bfk)=-t\sum_{j=1,2,3}\delta_j\cdot\hat{\bf
x}e^{i\bfk\cdot\delta_j}$. Neglecting vertex corrections, we calculate the d.c. conductance at zero temperature according to
\begin{align}
G_0=&
	\int
		|\Phi(\bfk)\,
		\im{G^r}(\bfk,\omega=0)|^2
        d{\bf k}/(2\pi)^2.
\end{align}
We obtain a monotonically increasing conductance as
the distortion of the lattice increases, see Fig. \ref{fig-model} (d), which supports the
conclusion that our two-dimensional amorphous carbon lattice is
metallic. The same conclusion is drawn from the approximate model
based on the dispersion relation in Eq. (\ref{eq-disp}).


In summary, we have performed first principles theory as well as
considered a model Hamiltonian, to demonstrate that graphene can
become metallic, once a significant disorder is introduced in the
material. In the present work, this has been achieved by a realistic 
amorphization in 3 dimensions, using the stochastic quenching method. 
Our results show a transition 
from the
complex electronic structure associated with perfectly crystalline
graphene
to a regular conducting behavior. 
This transition occurs rather drastically as a function
of the amount of disorder in the materials.
Experimentally, disorder and non-crystallinity is commonly obtained by
ion-irradiation of a crystalline material, or by rapid thin-film growth, and it
is likely that this is an experimental way forward also in this case. The energy
associated with some of the amorphous structures considered here is only 0.43
eV/atom higher than that of crystalline graphene.

If realized experimentally, a much desired metallic and transparent thin film
has been found, with the possibility to have impact in applications, e.g. in
touch screens. Given that other allotropes of C can become superconducting, at
least when doped, this may also be explored for amorphous graphene.

Calculations were performed on Ainil, the supercomputer at the physics
institute at UACH funded by Chilean FONDECYT projects 1110602 and 11080259.
Financial support by the EU-India FP-7
collaboration under MONAMI is acknowledged, as well as the Swedish Research Council (VR),
the foundation for strategic research (SSF), the European Research Council (ERC),
the KAW foundation and the Swedish National Infrastructure for
Computing (SNIC).
JF thanks Allan Gut for fruitful discussions.



\end{document}